\documentclass[aps,prb,superscriptaddress,showpacs,amsmath,twocolumn]{revtex4}

\usepackage{amsmath}
\usepackage{amssymb}
\usepackage{graphicx}

\begin{document}


\title{Role of the trigonal warping on the minimal conductivity of 
bilayer graphene}

\author{ J{\'o}zsef Cserti
}
\affiliation{Department of Physics of Complex Systems,
E{\"o}tv{\"o}s University
\\ H-1117 Budapest, P\'azm\'any P{\'e}ter s{\'e}t\'any 1/A, Hungary}

\author{Andr\'as Csord\'as
}
\affiliation{
HAS-ELTE, Statistical and Biological Physics Research Group, 
\\ H-1117 Budapest, P\'azm\'any P{\'e}ter s{\'e}t\'any 1/A, Hungary}

\author{Gyula D\'avid
}
\affiliation{Department of Atomic Physics,
E{\"o}tv{\"o}s University
\\ H-1117 Budapest, P\'azm\'any P{\'e}ter s{\'e}t\'any 1/A, Hungary}

\date{\today}


\begin{abstract}

Using a reformulated Kubo formula we calculate the zero-energy minimal
conductivity of bilayer graphene taking into account the small but finite 
trigonal warping.  
We find that the conductivity is independent of the strength of
the trigonal warping and it is three times as large 
as that without trigonal warping, and six times larger than 
that in single layer graphene.
Although the trigonal warping of the dispersion relation 
around the valleys in the Brillouin zone is effective only 
for low energy excitations, our result shows that its role cannot be neglected 
in the zero-energy  minimal conductivity. 

\end{abstract}


\maketitle


Recent experiments have proved that the charge carriers in graphene 
(single or stacks of atomic layer of graphite) are massless Dirac fermions
\cite{Novoselov-1:cikk,Kim:cikk,Novoselov-bilayer:cikk}. 
For recent reviews on graphene see
Refs.~\onlinecite{Katsnelson:rev,Katsnelson_Novoselov:rev,Geim_Novoselov:rev}.
Besides the unusual transport properties observed and reviewed in the
above works another important 
experimental feature is the minimal conductivity of the graphene
systems which was considered theoretically\cite{Fradkin_Lee_Gusynin-1:cikk} 
long before the experimental evidence.  
After the above mentioned experimental works on graphene, number of
theoretical studies\cite{Ziegler-robust:cikk,Falkovsky:cikk,Nomura_MacDonald:cikk,Gusynin-Hall_sxx:cikk,Gusynin-Dirac-transport:cikk,Peres-1:cikk,Katsnelson-s:cikk,Ostrovsky:cikk,Tworzydlo:cikk,sajat-bilayer:cikk,Ryu-Ludwig:cikk}
have predicted the conductivity of the order of $e^2/h$.  
Very recently, Miao et al.\ have experimentally 
confirmed\cite{Miao_exp_billiard:cikk} most theoretical 
predictions\cite{Gusynin-Hall_sxx:cikk,Peres-1:cikk,Katsnelson-s:cikk,Ostrovsky:cikk,Tworzydlo:cikk,sajat-bilayer:cikk,Ryu-Ludwig:cikk}, 
namely the minimum conductivity 
in wide and short strips approaches the universal
value $\sigma^{\text min}_{xx} = (4/\pi)\, e^2/h$ in single layer
graphene.

The bilayer graphene has been studied first 
experimentally\cite{Novoselov-bilayer:cikk} by Novoselov et al.\  and 
theoretically\cite{Ed-Falko:cikk} by McCann and  Fal'ko. 
McCann have calculated the asymmetry gap in the electronic band
structure of bilayer graphene\cite{Ed_asymmetry_gap}.
In biased bilayer graphene it was demonstrated that the gap can be
tuned by  electric field effect\cite{Castro_gap-tune:cikk}.
In bilayer graphene the semiconductor gap 
has recently been controlled experimentally 
by Ohta et al.~\cite{Ohta_bilayer:cikk}. 
The optical and magneto-optical far infrared properties of bilayer
graphene has been studied by Abergel et al.\cite{Abergel_Falko:cikk}.
The role of the impurities in biased bilayer graphene has been studied
by Nilsson and Neto\cite{Nilsson_Neto:cikk}. 
Ludwig has considered the conductance of a normal-superconductor
junction in bilayer graphene\cite{Ludwig:cikk}.  
Recently, Koshino and Ando have investigated the transport in bilayer
graphene in self-consistent Born approximation\cite{Ando-bilayer:cikk} 
and they found that in the strong-disorder regime 
$\sigma^{\text min}_{xx} = (8/\pi)\, e^2/h$, while in the weak-disorder
regime $\sigma^{\text min}_{xx} = (24/\pi)\, e^2/h$ which is six times
larger than in single layer graphene.
Similarly, Katsnelson has also calculated the minimal conductivity in
bilayers using the Landauer 
approach\cite{Katsnelson-bilayer:cikk} 
and he obtained a different value 
$\sigma^{\text min}_{xx} = 2\, e^2/h$.  
In Ref.~\onlinecite{sajat-bilayer:cikk} we found 
$\sigma^{\text min}_{xx} =  (8/\pi)\,e^2/h$ which was confirmed later 
by Snyman and Beenakker\cite{Snyman-Carlo:cikk} 
using the Landauer approach.

However, much fewer theoretical works paid attention to the role of 
the trigonal warping in bilayer graphene. 
The influence of the trigonal warping on the weak localization effect
has been investigated 
by Kechedzhi et al.~\cite{Kechedzhi_Falko_Ed_Altshuler:cikk}, while 
on the minimal conductivity only 
by Koshino and Ando\cite{Ando-bilayer:cikk} using an effective 
2 by 2 Hamiltonian. 
Our aim in this work is to calculate the minimal conductivity using the
Hamiltonian suggested originally by McCann and Fal'ko\cite{Ed-Falko:cikk}.
This Hamiltonian allows us to find the
zero-energy minimal conductivity as a function of the strength of 
the trigonal warping in bilayer graphene. 
We use the Kubo formula rewritten in a form
suitable for obtaining the zero-energy minimal conductivity 
in graphene systems.
Surprisingly, we find that the conductivity is {\em independent} of the 
strength of the trigonal warping and {\em six} times as large as that 
for single layer graphene.

The bilayer graphene consists of two coupled honeycomb lattices with
basis atoms $A_1, B_1$ and $A_2, B_2$ in the bottom and the top
layers, respectively. 
The two layers are arranged in Bernal stacking ($A_2-B_1$). 
The intralayer coupling between $A_1$ and $B_1$, and $A_2$ and $B_2$
is $\gamma_0$. 
The strongest interlayer coupling is between $A_2$ and $B_1$ with
coupling constant $\gamma_1$. 
A direct hopping between $A_1$ and $B_2$ is taken into
account by the coupling constant $\gamma_3 \ll \gamma_1$. 
This coupling is responsible for the trigonal warping. 
The above coupling constants are estimated as 
$\gamma_0 = 3.16$eV\cite{Toy-gamma_0:cikk}, 
$\gamma_1 = 0.39$eV\cite{Misu-gamma_1:cikk}, 
and $\gamma_3 = 0.315$eV\cite{Doezema-gamma_3:cikk}. 
 
To model the bilayer graphene we use the same gap-less Hamiltonian as
that in Ref.~\onlinecite{Ed-Falko:cikk} which takes into account the 
trigonal warping. 
The Hamiltonian in the basis $A_1, B_1, A_2, B_2$ in the valley 
${\bf K}$ and in the basis  $B_1, A_1, B_2, A_2$ in the valley 
${\bf K}^\prime$ reads 
\begin{eqnarray}
H_{b1} &=& \xi  \,  \left( \begin{array}{cccc}
0 & v p_- & 0 & v_3 p_+ \\
v p_+ & 0 & \xi \gamma_1 & 0 \\
0 & \xi \gamma_1 & 0 & v p_-  \\
v_3 p_- & 0 & v p_+ & 0  
\end{array}  \right), 
\label{Hfull:eq}
\end{eqnarray}%
where $p_\pm = p_x \pm i p_y$, 
$v = \sqrt{3}a \gamma_0/(2\hbar)$ and 
$v_3 = \sqrt{3}a \gamma_3/(2\hbar)$, while 
$\xi=+1$ for the valley ${\bf K}$ and $\xi=-1$  
for the valley ${\bf K}^\prime$ ($a=0.246$ nm is the lattice constant 
in the honeycomb lattice). 
The strength of the trigonal warping is desribed by the parameter
$\beta = v_3/v = \gamma_3/\gamma_0$. 
According to previous 
studies\cite{Toy-gamma_0:cikk,Doezema-gamma_3:cikk,Ed-Falko:cikk}
$\beta \approx 0.1$.
 
The four eigenvalues of the Hamiltonian (\ref{Hfull:eq}) 
as functions of the wave number 
${\bf k}= k(\cos\varphi,\sin\varphi)$ are given by 
\begin{eqnarray}
\lefteqn{
\hspace{-3mm} 
E_n^2(k,\varphi) = \frac{\gamma^2_1}{2}\left[
1+ \tilde{k}^2\left(\beta^2+2\right) +{\left(-1\right)}^n \Gamma \right], 
\, \text{where}
}   \\ 
\Gamma \!\! &=& \!\! \sqrt{1 \!\! - \!\! 2\tilde{k}^2\left(\beta^2-2\right)
\! + \! \tilde{k}^4 \beta^2 \left(\beta^2+4\right) 
\! + \! 8 \tilde{k}^3 \beta \cos 3\varphi
}, \nonumber 
\end{eqnarray}%
where $n=1,2$, while the rescaled wave number is 
$\tilde{k}= k\, \gamma_1/(\hbar v)$. 
\begin{figure}[hbt]
\includegraphics[scale=0.7]{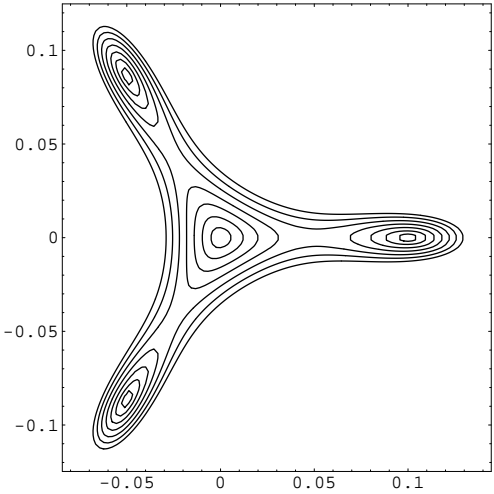}
\caption{\label{warping:fig}  
Constant energy lines (in units of $\gamma_1$) of the dispersion 
relation of the positive eigenvalue $E_1$ 
in the $(\tilde{k}_x,\tilde{k}_y)$ plane
around the ${\bf K}$ point of the Brillouin zone (at the origin in
this figure). 
Here $\beta = 0.1$ and the contour lines are plotted 
equidistantly with the most outer contour line corresponding 
to energy $2 E_L$.  
}
\end{figure}
Owing to the $\cos 3\varphi $ term the eigenvalues are 
three-fold rotational invariant for finite $\beta$. 
The eigenvalues $\pm E_1$ become zero at the ${\bf K}$ point of the
Brillouin zone, ie, at $\tilde{k}=0$,  and 
at the center of the three pockets located 
at $\tilde{k}=\beta$ and $\varphi = 0, 2\pi/3, 4\pi/3$. 
Around these zeros the constant energy lines are distorted as shown in 
Fig.~\ref{warping:fig}. 
This is called trigonal warping. 
At moderate energy, direct hopping between $A_1$ and $B_2$ leads 
to trigonal warping of the constant energy lines about each valley, 
but at an energy $E$ less than the Lifshitz energy 
$E_L = \gamma_{1} \beta^2/(4+\beta^2) \approx 1$meV, 
the effect of trigonal warping is dramatic. 
It leads to a Lifshitz transition\cite{Abrikosov:book}: 
the constant energy line is broken
into four pockets, which can be referred to as one central and three
leg parts. 
For $v_3 \ll v$, ie, $\beta \ll 1$ we find that the separation of the 2D Fermi
line into four pockets would take place for very small carrier
densities 
$n < n_{L} \sim 1\times 10^{11}\text{cm}^{-2}$. 
For $n<n_{L}$, the central part of the
Fermi surface at energy $E$ is approximately circular with area
$\mathcal{A}_{\mathrm{c}}\approx \pi E^{2}/(\hbar v_{3})^{2}$, and
each leg part is elliptical with area $\mathcal{A}_{\mathrm{\ell}}\approx
{\textstyle\frac{1}{3}}\mathcal{A}_{\mathrm{c}}$.
For $E\ll E_L $  the dispersion relation are linear in $k$. 
A similar structure of the constant energy lines can be seen around
the ${\bf K}^\prime$ point. 
For $\beta =0$ there is no trigonal warping, ie, the eigenvalues are
rotational symmetric and the Dirac cones are only at the ${\bf K}$ and
${\bf K}^\prime$ points. 

Recently, in self-consistent Born approximation 
Koshino and Ando\cite{Ando-bilayer:cikk} have investigated the minimal
conductivity for bilayer graphene using an approximated 2 by 2
Hamiltonian which mimics the trigonal warping and is given by   
\begin{eqnarray}
H_{b2} &=& 
g_2 \left( \begin{array}{cc}
0 & \tilde{p}_-^2 -  \tilde{p}_+ \\
\tilde{p}_+^2 - \tilde{p}_- & 0
\end{array}  \right) , 
\label{Ando-H:eq}
\end{eqnarray}%
where the effective coupling constant 
$g_2= \gamma_1 \gamma_3^2/\gamma_0^2$, 
the rescaled momentums are 
$\tilde{p}_\pm = (p_x \pm i p_y)/p_0$ 
and $p_0= 2\hbar \gamma_1 \gamma_3/(\sqrt{3}a \gamma^2_0)$.

The simplest effective Hamiltonian 
valid for $E\ll \gamma_1$ and first introduced 
by McCann and Fal'ko\cite{Ed-Falko:cikk} to study the Hall
conductivity of bilayer graphene is given by  
\begin{eqnarray}
H_{b3} &=& - g_3\, \left( \begin{array}{cc}
0 & p_+^2  \\
p_-^2  & 0
\end{array}  \right) , 
\label{Katsnelson-H:eq}
\end{eqnarray}%
where $g_3=v^2/\gamma_1$ is the effective coupling constant.
In this case the trigonal warping is absent. 
In this work, we calculate the minimal conductivity for all the three 
Hamiltonians, $H_{b1}, H_{b2}$ and $H_{b3}$.

To find the minimal conductivity for graphene systems we start from
the Kubo formula used by Ryu et al.~in Ref.~\onlinecite{Ryu-Ludwig:cikk}  
which at zero temperature and for dc conductivity 
(at zero frequency $\omega$) is given by
\begin{subequations} 
\begin{eqnarray}
\lefteqn{\sigma^{\text{min}}_{\mu\nu} = 
n_s n_v \lim_{\eta\to 0}\sigma_{\mu\nu} (\eta ), \,\,\, \text{where}
\label{sigma-main:eq}
}
\\ 
&& \hspace{-5mm}\sigma_{\mu\nu} (\eta ) =- \delta_{\mu\nu} \frac{\hbar}{4\pi}
\!\! \int \frac{d^2 {\bf r}}{S} 
\!\! \int d^2 {\bf r}^\prime
\Sigma_{\mu\nu}({\bf r},{\bf r}^\prime;E=0,\eta), 
\label{sigma-eta-1:eq}\\
&& \hspace{-5mm} \Sigma_{\mu\nu}({\bf r},{\bf r}^\prime;E,\eta)  =  
\text{Tr} \left[
G^{\text{A-R}}({\bf r},{\bf r}^\prime;E,\eta) j_\mu \right. \nonumber \\ 
&&\times  \left. G^{\text{A-R}}({\bf r},{\bf r}^\prime;E,\eta) j_\nu 
\right].
\end{eqnarray}%
Here $(\mu,\nu)=x,y$, the spin degeneracy
is $n_s=2$, the valley degeneracy
corresponding to the valley ${\bf K}$ and ${\bf K}^\prime$ is $n_v=2$, 
the area of the sample is $S$, while  
\begin{eqnarray}
G^{\text{A-R}}({\bf r},{\bf r}^\prime;E,\eta) 
\!\!\!\! &=&   \!\!\!\!\! G^-({\bf r},{\bf r}^\prime;E,\eta)
\! - \! G^+({\bf r},{\bf r}^\prime;E,\eta).
\end{eqnarray}%
The trace is taken over the spinor indices and 
for systems with translation invariance, the single-particle Green's
functions are given by 
\begin{eqnarray}
G^{\pm}({\bf r}_1,{\bf r}_2;E,\eta) &\!\!=& 
\!\!\!\! \int \frac{d^2 {\bf k}}{{\left(2\pi\right)}^2}\, 
e^{i {\bf k}\left({\bf r}_2-{\bf r}_1\right)} 
G^{\pm}({\bf k};E,\eta), \\
G^{\pm}({\bf k};E,\eta) &=& 
{\left[E\pm i \eta -H({\bf k}) \right]}^{-1},\\
H({\bf k}) &=& H({\bf p}= \hbar {\bf k}), 
\end{eqnarray}%
and the current operator is 
\begin{eqnarray}
j_\mu &=& i \frac{e}{\hbar}\, \left[H,r_\mu \right] = \frac{e}{\hbar}\,
\frac{\partial H({\bf k})}{\partial k_\mu}.
\label{current-main:eq}
\end{eqnarray}%
\end{subequations}%

The above expression (\ref{sigma-eta-1:eq}) 
for the conductivity can be simplified 
using the identity
\begin{equation}
\left(-z -H\right)^{-1} - \left(z -H\right)^{-1} = 
-2 z {\left(z^2 -H^2\right)}^{-1}. 
\end{equation}
Then with $z=i \eta$ we can rewrite $\sigma_{\mu\nu} (\eta )$ in 
Eq.~(\ref{sigma-eta-1:eq}) as 
\begin{subequations}%
\begin{eqnarray}
\sigma_{\mu\nu} (\eta ) &=& \delta_{\mu\nu}\,
\frac{2e^2}{h}\, \eta^2 I(\eta), \,\,\, \text{where}
\label{sigma-eta-2:eq}\\
I(\eta) &=&  \int \frac{d^2 {\bf k}}{{\left(2\pi\right)}^2}\, 
\text{Tr}\left[{\left[\eta^2+H^2({\bf k})\right]}^{-1} 
\frac{\partial H({\bf k})}{\partial k_\mu} \right. \nonumber \\
&\times& \left. 
{\left[\eta^2+H^2({\bf k})\right]}^{-1} 
\frac{\partial H({\bf k})}{\partial k_\nu}\right].
\label{int-sigma:eq} 
\end{eqnarray}%
\label{int-sigma-egybe:eq}
\end{subequations}%
This expression of the conductivity is valid for translational
invariant systems. 

Before we turn to the case of the bilayer it is instructive to see how
the expression (\ref{int-sigma:eq}) works for single layer graphene.   
In this case the Hamiltonian is given by 
\begin{equation}
H_s({\bf k})= g_s \left( \begin{array}{cc}
0 & k_-  \\
 k_+ & 0 
\end{array}  \right), 
\end{equation}
where $g_s = \hbar v$ and  $k_\pm = k_x \pm i k_y$.
The integrand in Eq.~(\ref{int-sigma:eq}) can easily be
calculated using the polar coordinates 
${\bf k}= k(\cos\varphi,\sin\varphi)$ and one finds    
\begin{eqnarray}
I(\eta) \equiv I_s(\eta) &=& \int_0^\infty \, 
\frac{2 g_s^2 k}{{\left(g_s^2 k^2+\eta^2\right)}^2}\,\frac{d k}{2\pi} 
= \frac{1}{2\pi  \eta^2},
\end{eqnarray}%
which is independent of the coupling constant $g_s$. 
Note that the main contribution in the integral $I(\eta)$ comes from the
vicinity of $k=0$, therefore the integral over $k$ 
can be extended to infinity\cite{Ostrovsky:cikk}.   
Then from  Eqs.~(\ref{sigma-eta-2:eq}) and (\ref{sigma-main:eq}) 
we obtain the well-known universal value of the minimal conductivity
for single layer graphene:  
$\sigma^{\text min}_{xx} = (4/\pi)\, (e^2/h)$
\cite{Gusynin-Hall_sxx:cikk,Peres-1:cikk,Katsnelson-s:cikk,Ostrovsky:cikk,Tworzydlo:cikk,sajat-bilayer:cikk,Ryu-Ludwig:cikk}.

We now consider the bilayer graphene taking into account the effect of
the trigonal warping. 
For bilayer with Hamiltonian (\ref{Hfull:eq}) 
the current operators defined by Eq.~(\ref{current-main:eq}) 
have a simple form  
\begin{subequations}
\begin{eqnarray}
j_x &=&  \xi \frac{e v}{\hbar} \left( \begin{array}{cccc}
0 & 1 & 0 & \beta  \\
1 & 0 & 0 & 0 \\
0 & 0 & 0 & 1  \\
\beta & 0 & 1 & 0  
\end{array}  \right), 
\label{jx:eq} \\[2ex]
j_y &=& i \xi \frac{e v}{\hbar} \left( \begin{array}{cccc}
0 & -1 & 0 & \beta  \\
1 & 0 & 0 & 0 \\
0 & 0 & 0 & -1  \\
-\beta & 0 & 1 & 0  
\end{array}  \right). 
\label{jy:eq}
\end{eqnarray}%
\end{subequations}%
The integral $I(\eta)\equiv I_{b1} (\beta,\eta)$ 
in Eq.~(\ref{int-sigma:eq}) will depend on $\beta$. 
It can be shown that $\sigma_{xx}=\sigma_{yy}$, therefore it is 
convenient to calculate $\sigma_{xx}=(\sigma_{xx}+\sigma_{yy})/2$. 
Using the polar coordinates ${\bf k}= k(\cos\varphi,\sin\varphi)$ 
and rescaling the variables $k$ and $\eta$ as 
$k\to k\, \hbar v/\gamma_1$ and $\eta \to \eta/\gamma_1$, 
a straightforward algebra yields for the case of $\xi = +1$ (valley
${\bf K}$): 
\begin{subequations}
\begin{eqnarray}
\lefteqn {I_{b1} (\beta,\eta) = \int_0^\infty  \int_0^{2\pi}
\!\! k\, \frac{A +B\cos 3\varphi }
{{\left(C+D\cos 3\varphi\right)}^2} \, \frac{d \varphi}{2\pi}\, 
\frac{d k}{2\pi},   } \\
&& A = k^4\left(2+5\beta^2\right) \nonumber \\
&& +\left(1+\eta^2\right)
\left[4 k^2 \left(1+\beta^2\right)+ 2\eta^2
+\beta^2\left(1+\eta^2\right)\right],  \\
&& B = 4k^3\beta ,\\
&& C = k^4+ \eta^4 +\eta^2 
+ k^2\left[2 \eta^2+\beta^2 \left(1+\eta^2 \right)\right],\\
&& D = -2k^3\beta . 
\end{eqnarray}%
\end{subequations}%
This expression has a three-fold rotational
symmetry as should be for trigonally warped bilayer graphene. 
It can be shown that for $\xi=-1$ (valley ${\bf K}^\prime$) we have
the same results. 
The conductivity is two-fold degenerate according to the valleys,
ie, $n_v=2$. 

The integral over $\varphi$ can be performed analytically 
and $I_{b1} (\beta,\eta) $ becomes 
\begin{eqnarray}
I_{b1} (\beta,\eta) &=& 2 \! \int_0^\infty  k\, 
\frac{AC -BD}{{\left(C^2-D^2\right)}^{3/2}}\, \frac{d k}{2\pi}.
\label{int-rad-fullbilayer:eq}
\end{eqnarray}%
Without trigonal warping, ie, for $\beta=0$ one finds  
\begin{eqnarray}
I_{b1} (\beta=0,\eta) &=& \nonumber \\ 
&& \hspace{-30mm}\int_0^\infty \, \frac{d k}{2\pi}\, 
\frac{4 k \left(k^4+\eta^4+\eta^2 +2 k^2 +2k^2\eta^2\right)}
{{\left(k^4+\eta^4+\eta^2+2k^2\eta^2\right)}^{2}} 
= \frac{1}{\pi \eta^2}.
\end{eqnarray}
Thus, using Eqs.~(\ref{sigma-eta-2:eq}) and (\ref{sigma-main:eq}) 
the minimal conductivity for bilayer graphene without trigonal warping
is $\sigma^{\text min}_{xx}(\beta=0) = (8/\pi)\, (e^2/h)$.
This result has been derived first in Ref.~{\onlinecite{sajat-bilayer:cikk}}
in a different way, 
and subsequently by Snyman and Beenakker in
Ref.~\onlinecite{Snyman-Carlo:cikk} using the Landauer approach.

After a tedious calculation the integral in
Eq.~(\ref{int-rad-fullbilayer:eq}) for finite value of $\beta$
can be performed yielding 
\begin{eqnarray}
I_{b1} (\beta,\eta) &=& \frac{1}{4\pi \eta^2}\, \left(12-\frac{127+145
  \beta^2 +38\beta^4}{\beta^6+\beta^4}\eta^2 \right) \nonumber \\
&& +O(\ln \eta)+ O(\eta^2) . 
\end{eqnarray}
Again, using Eqs.~(\ref{sigma-eta-2:eq}) and (\ref{sigma-main:eq}) 
we find a remarkable result, namely the minimal conductivity 
for bilayer graphene with trigonal warping takes a universal value   
$\sigma^{\text min}_{xx}(\beta) = (24/\pi)\, (e^2/h)$ 
independent of the stregth $\beta$ of the warping. 
This is our central result in this paper. 
This value is six times as large as the conductivity 
in single layer graphene.
It is suprising that $\sigma^{\text min}_{xx}(\beta)$ is not a 
continous function  around $\beta=0$.
Indeed, as we have seen 
$\sigma^{\text min}_{xx}(\beta=0)=(8/\pi)\, (e^2/h)$, 
while for {\em any finite} values of $\beta$ it is three times larger. 
This non-analitic behaviour of  $\sigma^{\text min}_{xx}(\beta)$ at
$\beta=0$ is a consequence of the fact that the minimal
conductivity results from the electronic dynamics in the limit of
zero density $n \to 0$. 
For any non-zero $\beta$, such density is {\em always} 
below the Lifshitz density $n<n_{L}$ where the 2D Fermi line
around each valley forms four separate pockets, whereas for $\beta = 0$, the
Lifshitz transition does not occur and there is always a single Fermi line
at each valley.

In the frame work of self-consistent Born approximation 
the same result was predicted by Koshino and Ando~\cite{Ando-bilayer:cikk} 
using the Hamiltonian $H_{b2}$ given by Eq.~(\ref{Ando-H:eq}). 
Note that in this Hamiltonian there is no adjustable parameter for the
strength of the trigonal warping like $\beta$ 
for Hamiltonian (\ref{Hfull:eq}). 
The effective coupling constant $g_2$ drops out in
Eq.~(\ref{int-sigma:eq}), therefore in this model the trigonal warping 
is built in without the possibility to change its strength. 
Using Eq.~(\ref{int-sigma-egybe:eq}) we repeat the calculation with the
Hamiltonian (\ref{Ando-H:eq}) and find  
\begin{equation}
I(\eta) \equiv I_{b2}(\eta) = \frac{1}{4\pi \eta^2}\, 
\left(12-127\eta^2 \right)+ O(\ln \eta)+ O(\eta^2). 
\end{equation}
Thus, the minimal conductivity takes the same universal value
$\sigma^{\text min}_{xx}(\beta) = (24/\pi)\, (e^2/h)$  as that for
the Hamiltonian $H_{b1}$ given by (\ref{Hfull:eq}).   

Finally, we calculate the minimal conductivity using the simplest
Hamiltonian $H_{b3}$ given by Eq.~(\ref{Katsnelson-H:eq}). 
Then  the integral in~(\ref{int-sigma:eq}) becomes
\begin{equation}
I(\eta) \equiv I_{b3}(\eta) = \int_0^\infty \, \frac{d k}{2\pi} \, 
\frac{8 g_3^2 k^3}{\eta^2 +g_3^2 k^4} = \frac{1}{\pi \eta^2}.  
\end{equation}
Thus, the minimal conductivity takes the same universal value
$\sigma^{\text min}_{xx} = (8/\pi)\, (e^2/h)$  as that for
Hamiltonian (\ref{Hfull:eq}) with $\beta=0$.   

In summary, we compared the minimal conductivity in bilayer graphene
obtained from three different effective Hamiltonians used in the
literature. 
We found that for the case when the trigonal warping is absent,  
the conductivity is {\em always} two times larger,  
while in the presence of trigonal warping it is {\em six times} 
larger than that for single layer graphene and is independent of the
strength of the warping .
Our universal results suggests that the conductivity has a topological
origin, which can be a further research topic in the future.

We gratefully acknowledge discussions
with C. W. J. Beenakker, E. McCann, and V. Fal'ko.
This work is supported partially 
by European Commission Contract No.~MRTN-CT-2003-504574 and the
Hungarian Science Foundation OTKA 046129.

\end{document}